\documentclass[sigconf]{acmart}
\renewcommand\footnotetextcopyrightpermission[1]{\footnote{
\copyright 2018 Copyright held by the owner/author(s), published under Creative Commons CC BY 4.0 License; \newline
Accepted at WSDM 2019, February 11--15 2019, Melbourne, Australia
\includegraphics[scale=0.8]{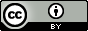}}} 
\usepackage{graphicx}
\usepackage{cleveref}
\usepackage{verbatim}
\usepackage{amsmath}
\usepackage{xfrac}
\usepackage{subfig}
\usepackage{balance}
\usepackage{tabularx}
\usepackage{url}
\usepackage{minibox}

\usepackage{xspace}

\usepackage{arabtex}
\usepackage{utf8}

\usepackage[hide]{chato-notes} 

\usepackage{cmm-greek}

\usepackage{longtable}

\usepackage{layouts}
\usepackage{textcomp}
\usepackage{dblfloatfix}
\usepackage{placeins}

\usepackage{microtype}

\usepackage{adjustbox}
\usepackage{array}
\usepackage{booktabs}
\usepackage{multirow}
\usepackage{tabularx}
\usepackage{enumitem}

\usepackage[normalem]{ulem}

\usepackage{tikz}
\usetikzlibrary{shapes,arrows,calc,positioning}

\usepackage{hyphenat}

\newcommand{\footnoteurl}[1]{\footnote{\url{#1}}}
\newcommand{\para}[1]{\smallskip\noindent\textbf{#1}}

\newcommand{\mytilde}{\raise.30ex\hbox{$\scriptstyle\mathtt{\sim}$}}

\newcommand{\ie}{\textit{i.e.}\xspace}
\newcommand{\eg}{\textit{e.g.}\xspace}
\newcommand{\cf}{\textit{cf.}\xspace}
\newcommand{\etal}{\textit{et al.}\xspace}
\newcommand{\vs}{\textit{vs.}\xspace}
\newcommand{\etc}{\textit{etc.}\xspace}

\newcommand{\Secref}[1]{Sec.~\ref{#1}}

\newcommand{\Tabref}[1]{Table~\ref{#1}}

\newcommand{\Figref}[1]{Fig.~\ref{#1}}

\newcommand\blfootnote[1]{%
  \begingroup
  \renewcommand\thefootnote{}\footnote{#1}%
  \addtocounter{footnote}{-1}%
  \endgroup
}

\newcommand{\authorbox}[3]{
\minibox[c]{
#1\\
{\fontsize{10pt}{10pt}\selectfont{}#2}\\
{\fontsize{10pt}{10pt}\selectfont{}#3}
}
}

\begin{document}

\settopmatter{printacmref=false, printfolios=false}





\newcommand{\myTitle}{Why the World Reads Wikipedia: Beyond English Speakers}
\title{\myTitle}


\copyrightyear{2019}
\acmYear{2019}
\acmConference[WSDM '19]{The Twelfth ACM International Conference on Web Search and Data Mining}{February 11--15, 2019}{Melbourne, VIC, Australia}
\acmBooktitle{The Twelfth ACM International Conference on Web Search and Data Mining (WSDM '19), February 11--15, 2019, Melbourne, VIC, Australia}
\acmPrice{15.00}
\setcopyright{acmlicensed} 
\acmDOI{...............} 
\acmISBN{.................} 

\author{
\authorbox{Florian Lemmerich}{RWTH Aachen University}{florian.lemmerich@cssh.rwth-aachen.de}
\authorbox{Diego S\'aez-Trumper}{Wikimedia Foundation}{diego@wikimedia.org}
\authorbox{Robert West*}{EPFL}{robert.west@epf\/l.ch}
\authorbox{Leila Zia}{Wikimedia Foundation}{leila@wikimedia.org}
}

\renewcommand{\shortauthors}{Florian Lemmerich, Diego S\'aez-Trumper, Robert West, and Leila Zia}



\begin{abstract}
As one of the Web's primary multilingual knowledge sources, Wikipedia is read by millions of people across the globe every day.
Despite this global readership, little is known about why users read Wikipedia's various language editions.
To bridge this gap, we conduct a comparative study by combining a large-scale survey of Wikipedia readers across 14 language editions with a log-based analysis of user activity.
We proceed in three steps.
First, we analyze the survey results to compare the prevalence of Wikipedia use cases across languages, discovering commonalities, but also substantial differences, among Wikipedia languages with respect to their usage.
Second, we match survey responses to the respondents' traces in Wikipedia's server logs to characterize behavioral patterns associated with specific use cases, finding that distinctive patterns consistently mark certain use cases across language editions.
Third, we show that certain Wikipedia use cases are more common in countries with certain socio\hyp economic characteristics; \emph{e.g.}, in\hyp depth reading of Wikipedia articles is substantially more common in countries with a low Human Development Index.
These findings advance our understanding of reader motivations and behaviors across Wikipedia languages and have implications for Wikipedia editors and developers of Wikipedia and other Web technologies.
\end{abstract}

\maketitle

\blfootnote{*Robert West is a Wikimedia Foundation Research Fellow.}




\section{Introduction}
\label{sec:intro}

Wikipedia is the world's largest encyclopedia and one of the primary knowledge sources on the Web, providing content every day to millions of readers from across the globe in more than 160 actively edited languages. Despite its global reach, very little is known about Wikipedia readers' motivations and information needs across languages. For years, English Wikipedia has been the primary focus of Wikipedia studies, and this has had implications on the way Wikipedia has been developed and supported over the years. In this study, we challenge the focus on English Wikipedia by expanding an earlier study \cite{singer2017we} in order to better understand the readers behind different Wikipedia languages. Without understanding similarities and differences between readers across the globe, improving user experience through new content, products, and services will continue to be challenging \cite{basu2003context,feild2010predicting}.

\para{Background and objectives.}
Most research on user motivation and needs has been dedicated to understanding the content producer perspective~\cite{nov2007motivates,arazy2017and}. Only recently, a study conducted on the English Wikipedia investigated why users read Wikipedia, via a large-scale user survey \cite{singer2017we}.
However, the focus on the English Wikipedia in that study neglects that, even under similar technical preconditions, the usage of Web contents can significantly differ depending on the cultural background of users \cite{qiu2013cultural,chen2011comparision}.
In contrast, the present work aims to understand \emph{why the world reads Wikipedia}.

\para{Materials and methods.} We base our analysis on a large-scale multiple-choice survey with questions identical to previous research \cite{singer2017we}, but with a massively extended scope---engaging readers of 14 Wikipedia languages and receiving more than 210,000 responses. 
Linking the survey participants to their traces in Wikipedia's server logs and comparing the data with the traces of random samples of readers allows for correcting misrepresentation of user groups and enables us to identify associations between usage patterns in the log data and specific use cases of Wikipedia that hold consistently across languages.
Furthermore, we employ country-level datasets to correlate Wikipedia's use cases with socio-economic and cultural indicators.

\para{Contributions and findings.} The following are our main contributions: 
(i)~We quantify and compare the prevalence of Wikipedia use cases with respect to motivations, information needs, and prior familiarity across 14 Wikipedia languages via a large-scale survey
(\Secref{sec:survey_results}).
(ii)~We match survey responses to the respondents' traces in Wikipedia's server logs to characterize usage patterns associated with specific use cases (\Secref{sec:result_logs}). 
(iii)~We match the survey data with country-level socio-economic and cultural data to allow for a deeper exploration of survey responses (\Secref{sec:result_country}).
 
Based on our analysis, we conclude that Wikipedia is read for a wide variety of use cases in any given language, and the distribution of use cases differs substantially between the languages. English Wikipedia is not fully representative of other Wikipedia languages. Additionally, we conclude that several (but not all) Wikipedia use cases can be associated with similar usage patterns across Wikipedia languages. Finally, we observe that socio-economic characteristics of a reader's country show remarkable correlations with the prevalence of Wikipedia use cases. 
For example, readers from less developed countries are more likely to be motivated by intrinsic learning and to read articles in depth.

The outcomes of this research can help Wikipedia editors across languages, Wikipedia developers, and the Wikimedia Foundation to create content and build tools and services with a deeper understanding of the needs of Wikipedia readers across the globe.

\section{Related Work}
\label{sec:related}

To understand Wikipedia readers across languages, our study draws on three different lines of research, described next. 

\para{Cultural differences in social media.}
Notable differences in the usage of social media platforms across countries and cultures have been found in a wide variety of platforms, such as Foursquare~\cite{silva2014you}, Yahoo!\ Answers~\cite{kayes2015cultures}, Twitter~\cite{garcia2013cultural,poblete2011all}, and Google+~\cite{magno2012new}. Moreover, previous studies show that, although Chinese sites like Renren and Weibo are technically very similar to Facebook and Twitter, their culture is perceived as more collectivist, suggesting that cultural background could be more important than the technology used in describing the observed usage differences \cite{qiu2013cultural,chen2011comparision}. A comprehensive survey on HCI and cultural differences \cite{kyriakoullis2016culture} emphasizes that understanding cultural values is essential for the design of successful user interfaces. Certain aspects of social media usage, \emph{e.g.}, topics discussed, could also be linked to socio-economic factors in Foursquare \cite{quercia2014mining,venerandi2015measuring} and Twitter \cite{quercia2012tracking}.

\para{Wikipedia across countries and languages.}
Several independent studies have covered specific aspects of cultural differences on Wikipedia. It was found that Wikipedia language editions have a high degree of self-focus, \emph{i.e.}, bias towards the knowledge of the editor community \cite{hecht2009measuring,miquel2018wikipedia} and set different priorities on the information included \cite{callahan2011cultural,jiang2017wikipedia,laufer2015mining,Samoilenko2017History}.
Those studies all focus on the editor or content perspective of Wikipedia, while in this paper we investigate the motivations and behaviors of readers.

\para{Wikipedia users' behavior and motivations.}
The behavior of Wikipedia readers has also been a main topic of interest, but primarily focused on content popularity~\cite{lehmann2014reader,ratkiewicz2010characterizing,spoerri2007popular} and navigation patterns~\cite{west2012human,singer2014detecting}. 
Studies of the motivations of Wikipedia users focused mainly on contributors \cite{nov2007motivates,arazy2017and}. By contrast, the motivation of readers has only been picked up recently in the predecessor study of this work, which studied reader motivation in the English Wikipedia only~\cite{singer2017we}.
All these studies neglect the multilingual and cross-cultural perspective that is the focus of this paper.

\section{Datasets and Methodology}
First, we describe the datasets used in this work in detail.
\subsection{Survey Data}
We selected 14 Wikipedia languages (\cf \Tabref{tab:languages}) with the following considerations:
the language family,
language\hyp specific Wikipedia article and pageviews counts,
and the number and distribution of speakers worldwide.
We also took into account the requests by Wikipedia volunteers to include their languages as part of the study.

\begin{table}[t]
\caption{The surveyed Wikipedia languages with the number of articles, the number of pageviews in the survey period, the sampling rate used for selecting survey participants, and the number of responses.}
\label{tab:languages}
\small
\begin{tabular}{ll|rr|rr}
\toprule
language & lang &   \# articles &  \# pageviews  & rate & \# resp. \\
\midrule
Arabic & ar &    523,917 &    38,102,782 & 1:10 & 2,158\\
Bengali &  bn &     51,015 &     1,865,887 & 1:1 & 1,198\\
German &  de &   2,079,460 &   227,823,185 & 1:5 & 28,000\\
English &  en &   5,414,505 &  1,945,323,873 & 1:40 & 24,140\\
Spanish &  es &   1,292,245 &   264,464,604 & 1:5 & 39,021\\
Hebrew &  he &    208,859 &    14,088,014 & 1:3 & 8,848 \\
Hindi &  hi &    121,867 &     9,041,447 & 1:2 & 3,064\\
Hungarian &  hu &    412,483 &    11,436,690 & 1:2.5 & 2,455\\
Japanese &  ja &   1,065,498 &   307,436,312 & 1:5 & 19,996\\
Dutch &  nl &   1,904,240 &    43,017,893 & 1:8 & 3,277\\
Romanian &  ro &    377,090 &     8,302,363 & 1:2 & 3,829\\
Russian &  ru &   1,402,293 &   224,732,227 & 1:5 & 67,621\\
Ukrainian &  uk &    703,665 &    12,446,880 & 1:2.5 & 8,041\\
Chinese &  zh &    946,356 &   116,703,091 & 1:20 & 5,957\\
\bottomrule
\end{tabular}
\end{table}

The survey was run from June 22 to June 29, 2017. The sampling rates were chosen with the intention to obtain roughly 30,000 responses from high\hyp pageview languages \vs\ 3,000 from the lower\hyp pageview languages, resulting in sampling rates ranging from 1:40 for English Wikipedia to 1:1 for Bengali Wikipedia (\Tabref{tab:languages}). We sampled from all users with requests to the specific Wikipedia languages' mobile and desktop sites, excluding requests to non\hyp article pages (discussion pages, search pages, \etc), those to the main page of Wikipedia and from browsers with \emph{Do Not Track} enabled. Potential survey participants were marked by assigning a token to their browsers. They were then shown a survey widget inviting them to participate in a three\hyp question survey to improve Wikipedia. The reader had the choice to ignore the message, dismiss it, or opt in to participate. This would take the reader to an external site (Google Forms) with a questionnaire titled \emph{``Why are you reading this article today?''} that contained, in random order, the following questions on
their \emph{motivation}, \emph{information need}, and \emph{prior knowledge}, respectively:
\begin{itemize}[leftmargin=*, topsep=1pt,after=\vspace{1pt},itemsep=-1ex,partopsep=1ex,parsep=1ex]
\item \emph{I am reading this article because\dots}: I have a work- or school-related assignment; I need to make a personal decision based on this topic (\eg, buy a  book, choose a travel destination); I want to know more about a current event (\eg, a soccer game, a recent earthquake, somebody's death); the topic was referenced in a piece of media (\eg, TV, radio, article, film, book); the topic came up in a conversation; I am bored or randomly exploring Wikipedia for fun; this topic is important to me, and I want to learn more about it (\eg, to learn about a culture); other. 
Users could select multiple answers for this question.
\item \emph{I am reading this article to\dots}: look up a specific fact or get a quick answer; get an overview of the topic; get an in-depth understanding of the topic. 
\item \emph{Prior to visiting this article\dots}: I was already familiar with the topic; I was not familiar with the topic, and I am learning about it for the first time. 
\end{itemize}

Prior to submitting their survey answers, readers were informed through a privacy statement\footnote{\url{https://wikimediafoundation.org/wiki/Survey_Privacy_Statement_for_Schema_Revision_15266417}} about
the collection, sharing, and usage of the survey data. Translations of the questions, answers, and privacy statement were provided by known Wikipedia editors and in close collaboration with one of the study authors, to preserve the specifics in the translated texts (\cf\ also \Secref{sec:limitations}).

We obtained more than 210,000 survey responses after removing empty or incomplete responses as well as responses that could not be mapped to a user trace (\cf\ \Secref{sec:auxdata}). \Tabref{tab:languages} displays the breakdown of responses by language. 
The survey responses along with the associated article information are made publicly available along with extended results from this paper.\footnote{\url{https://meta.wikimedia.org/wiki/Research:Characterizing_Wikipedia_Reader_Behaviour/Data}}

\subsection{Auxiliary Data}
\label{sec:auxdata}

We are interested in understanding how users' motivation, desired depth of knowledge, and prior knowledge (\ie, their answers to our survey) are reflected in reading behavior across languages and whether they can be explained through the socio-economic and the cultural context the users operate in. For this purpose, we link survey responses to the auxiliary data sources described here.

\para{Webrequest logs and article data.}
To analyze respondents' reading behavior in context and to apply bias correction to the survey data collected, we connect survey responses to Wikipedia's webrequest logs. For every request, the corresponding webrequest log contains, among others, the referrer URL, timestamp, client IP address, browser version, and rough geo-location derived from the client IP. Due to the absence of unique user IDs in the webrequest logs, we rely on the concatenation of client IP and 
user agent
as a pseudo ID. 
To obtain additional information on the requested articles, we extract the text of all articles and the Wikipedia link network for the 14 languages from the Wikipedia dumps\footnote{\url{https://dumps.wikimedia.org/}} of July 2017 (the dump following the survey period).
We then follow the methodology of Singer \etal\ \cite{singer2017we} to construct sessions for each user ID and extract a variety of features for each webrequest log entry. These features include
\begin{itemize}
[leftmargin=*, topsep=1pt,after=\vspace{1pt},itemsep=-1ex,partopsep=1ex,parsep=1ex]
    \item \emph{request features} such as the  country or continent of the user, local time, requested Wikipedia host (mobile or desktop), and referrer type (internal navigation, external search engine, or other);
    \item \emph{article features} such as the degree in the link network for this Wikipedia language, PageRank, text length, and topic (derived via Latent Dirichlet Allocation~\cite{blei2003latent} with $n=20$ topics separately for each language);
    \item \emph{activity features} such as the number of articles requested, duration of the session in minutes, time between two requests, and number of sessions during the survey period.
\end{itemize}

In addition to the survey participants' webrequest logs, we also select a fully random sample of 200,000 Wikipedia readers per language and compute the same set of features for them to enable bias correction (\Secref{sec:bias_correction}).

\para{Country-level data.}
For a more detailed analysis of the survey responses in the context that survey respondents are in, we connect the survey data and webrequest logs with two external datasets: first, the Quality of Government dataset~\cite{qog_standard_jan17}, which provides rich information on a large range of socio-economic statistics at the country level; and second, the well-known Hofstede dimensions~\cite{hofstede1984hofstede,hofstede1991cultures}, which describe the culture and values within a society.\footnote{\url{https://geerthofstede.com/research-and-vsm/dimension-data-matrix/}}

\subsection{Correcting Survey Bias}
\label{sec:bias_correction}
Inferring properties of a general population from a research survey is subject to different kinds of biases, including \emph{coverage bias}, \emph{sampling bias}, and \emph{non-response bias}. To correct for non-response bias, we apply a weighting scheme that gives higher weights to survey participants with user features that are underrepresented in the set of survey participants compared to the representative random sample.
For this purpose, we use inverse propensity score weighting~\cite{lunceford2004stratification,austin2011introduction} based on a gradient boosting classifier, as described in detail by Singer \etal\ \cite{singer2017we}. We calculate response weights for each language version independently.

\section{Results}
This section describes results on why users across the world read Wikipedia articles.

\subsection{Survey Results}
\label{sec:survey_results}
\todo{[Leila] Manually improve the plots if there's enough time or for camera ready.

{\color{red}[fle]} Please add suggestions. I spend yesterday already two hours making these plots more readable and taking less space. Will probably do this for the final version, though.
}

\begin{figure*}[t!]
	\centering
	\subfloat[Motivation]{\includegraphics[width=0.95\textwidth]{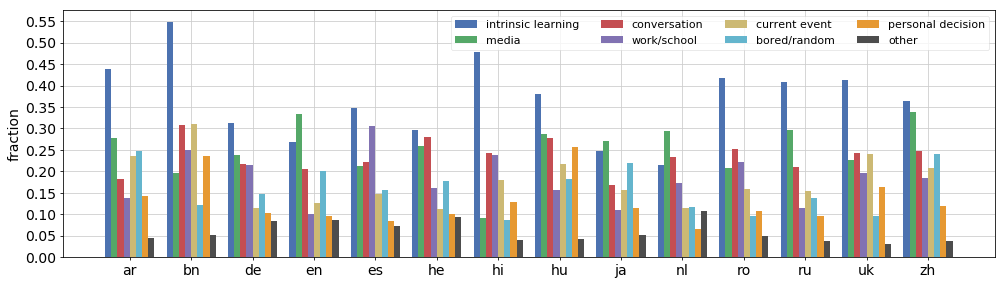} 
	\label{fig:survey_motivation_by_lang}}
	
	\subfloat[Information need]{	\includegraphics[width=0.49\textwidth]{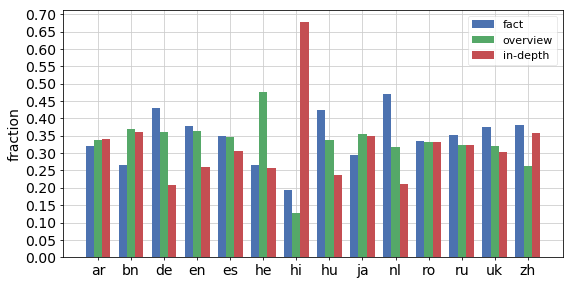} 
	\label{fig:survey_infoneed_by_lang}}
	\subfloat[Prior knowledge]{	\includegraphics[width=0.49\textwidth]{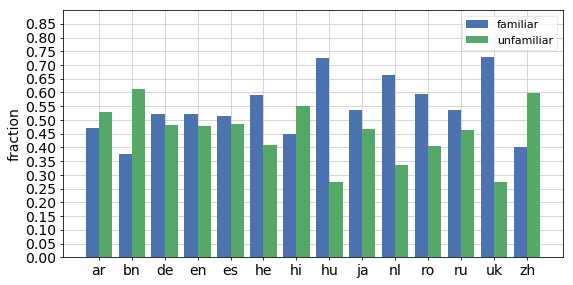} 
	\label{fig:survey_priorknowledge_by_lang}}

	\caption{Each group of bars reflects the responses in one language edition, each bar represents the share of one response option. For motivation, multiple answer options were allowed for each user.
	}
	\label{fig:survey_by_lang}
\end{figure*}

We start by presenting the distribution of responses to the survey questions across the 14 Wikipedia languages. We compute the weighted percentages of survey respondents with specific motivations, information needs, and prior knowledge, where the weights were computed as described in \Secref{sec:bias_correction}. We visualize the results in \Figref{fig:survey_by_lang}.

\para{Motivation.} 
\note{
\begin{itemize}
    \item we want to say a few main things in this section? Shall we try to condense the results and let the reader read up the tables if they wish to?  {\color{red}[fle:]I think its ok to spend some space discussing particular findings.}
    The things we want to mention: 1) There is a range of motivations, no dominant one, for all languages. 2) For all except 3 languages intrinsic learning is the lead motivator, 3) media is important, too, especially for the 3 languages where intrinsic learning is not the lead. 4) There are some major differences between some of the languages re bored/random and work/school. 5) only a small percentage have chosen "other" as an option, which says the taxonomy identified in the earlier work is robust. 6) We can't keep the reader hanging. We need to provide a glimpse into what can potentially explain the major differences explained in the sections after, with reference to those sections.
    {\color{red}[fle]: That makes for a great transition to the next Section.}
\end{itemize}
}
With respect to motivation (\Figref{fig:survey_motivation_by_lang}), we observe that Wikipedia is read with a wide range of motivations. In none of the languages does a single motivation clearly dominate. 
\emph{Intrinsic learning} is the most commonly selected motivation (mean: 37\%\footnote{Note that this is the unweighted mean of the outcomes for the surveyed language editions. Weighting by language edition size would neglect small editions.}) across all but three languages: English, Dutch, Japanese. For these three languages, \emph{media} is the top reported motivation instead (mean: 25\%), which is also one of the top motivations for all other languages with the exception of Bengali and Hindi. 
We also observe that considering \emph{intrinsic learning}, there are major differences between language editions: the response shares are generally lower for Western European languages (Dutch:~21\%, English:~27\%), and substantially higher for Eastern European languages (Romanian:~42\%, Russian:~41\%, Ukrainian: 41\%) as well as
Arabic (40\%) and Indian languages (Bengali:~55\%, Hindi:~48\%). 
Other common motivators are \emph{conversations} (mean: 24\%), \emph{work- or school}\hyp related tasks (mean: 18\%), \emph{current events} (mean: 17\%), and the need for making \emph{personal decisions} (mean: 13\%). 
Finally, we observe that the percentage of respondents being motivated by \emph{work- or school}\hyp related tasks or \emph{being bored} differs significantly across languages. While work- or school\hyp related motivations account for 10\% in English Wikipedia, they account for over three times as much (31\%) in Spanish Wikipedia.
Also, people report being bored as a motivation for visiting Wikipedia in only 10\% of responses in Hindi, Romanian, and Ukrainian Wikipedia, and in more than 20\% of responses in English, Japanese, Chinese, and Arabic Wikipedia.
We further note that the answer \emph{``other''} was selected only rarely (at most 10\%), indicating the robustness of the taxonomy defined in earlier research \cite{singer2017we}.

\para{Information need.} Considering the information need of readers, we observe that, considering all languages, Wikipedia is visited roughly equally by readers for in-depth understanding (mean: 32\%), fact checking (mean: 35\%), and obtaining an overview (mean: 33\%). We find, however, much diversity between languages. In-depth reading is reported substantially less often for the Western and Central European languages such as English (26\%), German (21\%), Hungarian (24\%), or Dutch (21\%). Instead, Wikipedia is more often used for fact checking in these language versions (38\%, 43\%, 43\%, and 47\%, respectively).
An outlier is the Hindi language, where users report in-depth reading of articles 68\% of the time. In \Secref{sec:result_country}, we explore this further in the light of socio-economic factors.

\para{Prior knowledge.} There are nearly the same numbers of people reporting to be familiar \vs\ unfamiliar with the topic they read on Wikipedia across languages (55\% \vs\ 45\%). This being said, there are substantial differences between the languages: Eastern European languages report familiarity with the content at much higher rates (Ukrainian: 73\%, Hungarian: 73\%), while Asian languages with the exception of Japanese report to be unfamiliar more often (Bengali: 61\%, Chinese: 60\%, Hindi: 55\%). These differences could potentially be  explained by a tradition and social desirability of humility in these Asian societies~\cite{monkhouse2013measuring} (\cf\ \Secref{sec:discussion}).

\begin{figure}[t!]
	\centering
	\includegraphics[width=0.99\columnwidth]{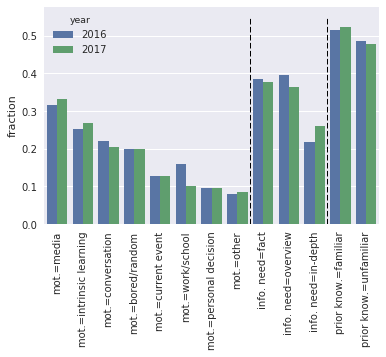}
	\caption{Comparison of English Wikipedia surveys for 2016 (blue) \cite{singer2017we} and 2017 (green). We can observe that overall survey responses are stable. Only the share of answers for \emph{motivation = work/school} decreased noticeably.}
	\label{fig:2016 vs 2017}
\end{figure}

\para{Robustness over time.} We examine the reproducibility and stability of the survey results over time by comparing the prevalence of English Wikipedia use cases with the results of the earlier study conducted on English Wikipedia in 2016 \cite{singer2017we}.
\Figref{fig:2016 vs 2017} shows that the survey results are very similar, suggesting that the observed effects are robust. The only noticeable difference between the results is a decrease in work- or school\hyp related motivation (16\% in March 2016 \vs\ 10\% in June 2017), which may be due to seasonal effects. 
\subsection{Survey Results and Webrequest Logs}
\label{sec:result_logs}
In the previous study on use cases in the English Wikipedia, certain use cases (motivation, information need, and prior knowledge) could be linked to specific usage patterns extracted from Wikipedia's server logs (webrequest logs)~\cite{singer2017we}.
In this section, we investigate whether such patterns are common across languages or whether they are only present in a subset of the languages.

We start by manually extracting binary usage patterns (such as $session\_length \geq 3$) from the log request features based on the previous study results \cite{singer2017we}, which have been detected using pattern mining techniques~\cite{kloessgen1996}.
The binarization allows for applying a single framework for categorical and continuous features with vastly different distributions.
For each language and each usage pattern, we can then compute the \emph{share} $S$ of users for which the pattern applies (\emph{e.g.}, a share of 20\% of the users have a $session\_length \geq 3$).
Additionally, we can compute for each language the \emph{effect} $E$ of any survey question answer on any pattern, \emph{i.e.}, the difference between the percentage of users for which the pattern applies among users that gave a specific survey response and the percentage of users for which the pattern applies among users that gave a different response to the respective survey question.
For example, if for a language the share of users with a $session\_length \geq 3$ is 25\% for users with the motivation \textit{work/school,} and 20\% for users with another motivation, then the effect of $motivation = work/school$ on $session\_length \geq 3$ is 5\%.

To find interesting relationships, we investigate all 247 pairs of binarized usage patterns and survey responses.
Given the 14 languages of the survey, we obtain for each pair a distribution of those 14 shares and effects.
For summarization, we calculate then the mean share $\mu(S)$ of all languages as well as the relative standard deviation $rs(S)$ (standard deviation divided by the mean share, also known as coefficient of variation) as a measure of variability between the languages.
Furthermore, we compute the mean effect $\mu(E)$ of all languages, the relative mean effect (the mean effect divided by the mean share) $\bar{\mu}(E)$, the standard deviation $\sigma(E)$ of the effect, and the normalized standard deviation $\bar{\sigma}(E)$ of the effect (the standard deviation of the effect divided by the mean share).
Given these statistics, we are then most interested in pairs with a large (relative) mean effect, since these exhibit a strong dependency between use case and usage pattern across languages.
Among those pairs, we can then differentiate between the more general (\emph{i.e.}, consistent between language editions) dependencies, which exhibit a low standard deviation of the effect, and correlations that are more specific to certain languages, which exhibit a high standard deviation of the effect.

\begin{figure*}[t!]
	\centering
	\subfloat[Effect of motiv. = bored/random  \protect\\  on internal referrer]
	{\includegraphics[width=0.24\textwidth]{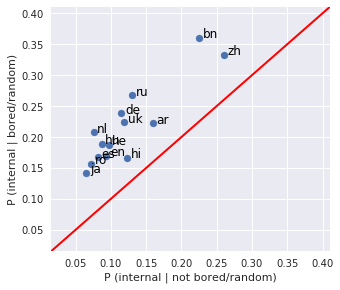}}
	\subfloat[Effect of motiv. = work/school \protect\\  on slow requests]
    {\includegraphics[width=0.24\textwidth]{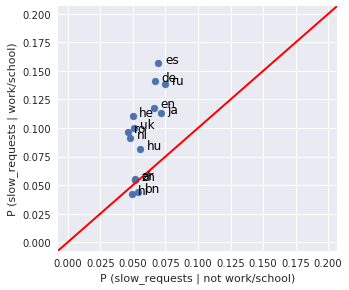}} 
	\subfloat[Effect of info. need = in-depth \protect\\ on long sessions] 
	{\includegraphics[width=0.24\textwidth]{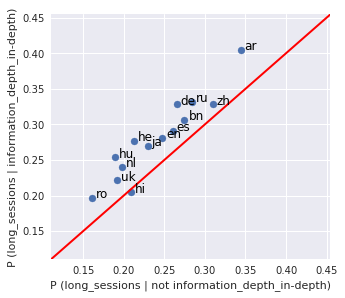}} 
	\subfloat[Effect of motiv. = current event \protect\\  on long articles] {\includegraphics[width=0.24\textwidth]{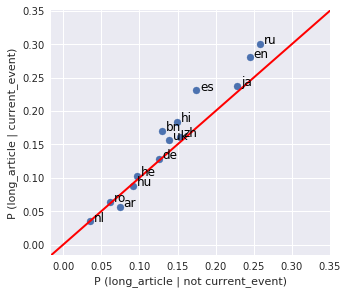}} 
	
	\caption{\textbf{Visualization of relationships between selected usage patterns and survey answers. Each plot shows one data point for each language, which indicates the likelihood of a usage pattern among users with a given survey response ($y$-axis) and among user that did not answer that way ($x$-axis). The diagonal is marked with a red line. This implies that for languages above the read line, answering with this survey response is more likely under this usage pattern.  }}
	\label{fig:rl}
\end{figure*}

We can visualize the relationship between a usage pattern and survey response across languages in plots such as shown in~\Figref{fig:rl}.
These plots display one point for each language edition. The coordinates of the point mark the probability of the usage pattern given a specific survey answer on the $y$-axis and the probability of the usage pattern given a different survey answer for this question on the $x$-axis.
Positive effects of the survey answer on the usage pattern are then indicated by points above the diagonal (\emph{e.g.}, all points in~\Figref{fig:rl}a), negative effects by points below the diagonal. The further the point for a language is from the diagonal, the stronger is the effect of the respective survey response on the usage pattern.

\begin{table}[b]
\small
\setlength\tabcolsep{0.5mm}
\caption{Pairs of usage patterns and survey responses with the largest normalized mean effect $\bar{\mu}(E)$ across language editions. This table provides for each pair information on the mean share (likelihood of the pattern) $\mu(S)$, and the relative standard deviation of the share $rs(S)$ across language editions. Furthermore it displays the mean effect (increase of the pattern likelihood in presence of the response) $\mu(E)$, the normalized mean effect $\bar{\mu}(E)$, the standard deviation $\sigma(E)$ and the normalized standard deviation of the effect $\bar{\sigma}(E)$.}
\label{tab:usage_patterns}
\begin{tabular}{llrrrrrrr}
\toprule
Pattern\footnotemark &  Response &  $\mu(S)$ &    $rs(S)$ &  $\mu(E)$ &  $\bar{\mu}(E)$ &  $\sigma(E)$  &  $\bar{\sigma}(E)$ \\
\midrule
       internal &        mot.=bored/rand. &        .136 &       .416 &         .095 &              .697 &        .028 &             .206 \\
  slow\_requests
 &         mot.=work/school &        .065 &       .220 &         .038 &              .594 &        .030 &             .457 \\
        desktop &         mot.=work/school &        .342 &       .303 &         .187 &              .547 &        .122 &             .358 \\
 rapid\_requests &        mot.=bored/rand. &        .102 &      393 &         .041 &              .405 &        .023 &             .229 \\
  long\_sessions &        mot.=bored/rand. &        .252 &           .204 &         .097 &              .383 &        .047 &             .188 \\
     time:night &        mot.=bored/rand. &        .112 &         .541 &         .031 &              .281 &        .032 &             .289 \\
   long\_article &        prior knowl.=familiar &   .143 &         .473 &         .036 &              .251 &        .032 &             .221 \\
 time:afternoon &         mot.=work/school &        .308 &           .116 &         .064 &              .207 &        .044 &             .142 \\
     time:night &       mot.=media &        .112 &           .541 &         .022 &              .197 &        .031 &             .281 \\
       internal &  mot.=intrinsic learn. &        .136 &            .416 &         .022 &              .163 &        .018 &             .131 \\
  long\_sessions &     info. need=in-depth &        .252 &            .204 &         .040 &              .158 &        .019 &             .075 \\
  slow\_requests &      mot.=other &        .065 &            .220 &         .009 &              .140 &        .021 &             .324 \\
     time:night &  mot.=intrinsic learn. &        .112 &            .541 &         .015 &              .131 &        .013 &             .114 \\
 weekday:Friday &     mot.=bored/rand. &        .113 &          .238 &         .015 &              .131 &        .018 &             .155 \\
       internal &     info. need=in-depth &        .136 &            .416 &         .017 &              .127 &        .015 &             .112 \\
  long\_sessions &        prior knowl.=familiar &        .252 &           .204 &         .032 &              .126 &        .022 &             .088 \\
        desktop &        mot.=other &        .342 &           .303 &         .042 &              .124 &        .058 &             .169 \\
  long\_sessions &  mot.=intrinsic learn. &        .252 &         .204 &         .030 &              .119 &        .024 &             .094 \\
     time:night &       prior knowl.=familiar &        .112 &         .541 &         .013 &              .118 &        .021 &             .192 \\
   long\_article &       mot.=current\_event &        .143 &        .473 &         .017 &              .117 &        .021 &             .144 \\
\bottomrule
\end{tabular}

\end{table}

\Tabref{tab:usage_patterns} shows the top pairs sorted by the relative mean effect.
For example, the top pattern (visualized also in~\Figref{fig:rl}a) shows that on average across languages  13.6\% of the users arrived on the surveyed page with an internal referrer, \emph{i.e.}, probably by browsing Wikipedia through links~\cite{dimitrov2017makes}. If according to the survey the user is bored or is randomly exploring Wikipedia, then the likelihood of an internal referrer is strongly increased, on average by 9.5\% percentage points or by 69.7\%.
The high relative standard deviation ($0.416$) indicates strong deviations between the languages.
\footnotetext{Slow and rapid requests are defined based on average amount of time between requests. Slow requests have on average more than 10 minutes in between while rapid request have on average less than 1 minute between requests;
Night time describes requests at a local time between midnight and 6 a.m., afternoon between noon and 6 p.m.; long session means at least 3 requests within the survey session; long article denotes articles with at least 40,000 characters.}%

We can find various patterns that are mostly consistent across language editions:
\begin{itemize}
[leftmargin=*, topsep=1pt,after=\vspace{1pt},itemsep=-1ex,partopsep=1ex,parsep=1ex]
    \item Users who randomly browse Wikipedia or are bored use internal navigation more often, browse many articles in one session, but do not stay long at individual articles, and view Wikipedia at night.
    \item Users with  work- or school\hyp related tasks have longer dwell times on the requested article, use Wikipedia's desktop version more often, and are more likely to visit Wikipedia in the afternoon.
    \item Similar to bored users, readers motivated by intrinsic learning more often have long sessions and browse at night times using internal navigation.
    \item By contrast, users who are motivated by a conversation or use Wikipedia for fact checking do so more often using the mobile platform, have shorter dwell times on articles, and use internal navigation less often.
    \item Users already familiar with a topic have longer sessions and request longer articles.
\end{itemize}
Most of the above\hyp mentioned dependencies are relatively consistent across language editions, as indicated by a small (normalized) standard deviation of the effect (less than 0.2). 
This indicates that, independent of the language, certain motivations correlate consistently with certain changes in usage behavior.
There are, however, also some specific exceptions. 
In particular, the effects of work- or school\hyp related motivations appear to differ between language editions (\Figref{fig:rl}b).

We also notice that, for most of the pairs, the average variability (as measured by the relative standard deviation) between the languages is much higher than the effects of survey responses (\cf\ \Figref{fig:rl}c and d for typical examples).
This observation implies that differences in the use cases alone are insufficient to explain the diverse prevalence of the usage patterns across languages.

Many noticeable correlations could already be observed in the initial study on the English language. However, not all dependencies found in the English language edition also hold in other languages.
For example, it was noted previously that users motivated by current events tend to read longer articles. While this effect is observed in the current survey in the English Wikipedia, it does not hold true considering all language editions (\Figref{fig:rl}d).

\subsection{Survey Responses and Country Statistics}
\label{sec:result_country}
Finally, we analyze if specific Wikipedia use cases can be associated with the socio-economic or cultural background of users in order to seek potential explanations for the differences between language editions.
To link available data for these factors to our survey results, we perform these analyses on a country level.

\para{Correlation of survey responses with socio-economic indicators.}
We start by correlating survey responses with socio-economic information.
In particular, we rely on the \emph{Human Development Index} (HDI), a summary statistic that reflects the development status of a country via the population and its capabilities and not only based on economic growth. It is the geometric mean of normalized indices defined under three dimensions: life expectancy, education, and income. The HDI is published by the United Nations Development Programme (UNDP) and is contained in the Quality of Government dataset (\Secref{sec:auxdata}).
Additionally, we also use Gross Domestic Product (GDP) per capita as well as the percentage of adults with secondary education as two other measures of country development.
Since a single Wikipedia language can be viewed in multiple countries and the HDI is reported at the country level, we partition the survey responses for each Wikipedia language by country.
For each language\slash country pair with at least 500 survey responses, we then compute the share of each answer option (\emph{e.g.}, \emph{motivation=media})  
for survey participants from this country and add country-specific statistics. Through this step, we obtain 43 language\slash country pairs, which we use as data points for a correlation analysis.

\Tabref{tab:corr_hdi} shows the Spearman correlation coefficient and its associated $p$-value (with Bonferroni correction for $n=13$ survey responses, but no correction for multiple attributes) when capturing the correlation between survey responses and HDI. We observe that several survey answers show significant correlations with HDI. 
For example, the more developed the country of a reader, the more likely the reader is to be motivated by  \emph{media} when visiting Wikipedia.
By contrast, being motivated by \emph{work or school} or by \emph{intrinsic learning} is more likely in developing or newly industrialized countries.
Regarding the information need of viewers, we can see that in-depth reading is more often reported in less developed countries, while in industrialized countries fact checking is a more prevalent use case.
Finally, reporting familiarity with a topic is also somewhat more common in industrialized countries.
We can conclude that there is a clear tendency towards in-depth reading and learning in less developed countries.

By zooming in on individual correlations, \emph{e.g.}, between HDI and \emph{intrinsic learning} (\Figref{fig:hdi_vs_intrinsic}), we obtain additional insights. We observe that in industrialized countries (United States, the Netherlands, Germany), linguistic minorities (such as Spanish speakers in the United States or Russian Speakers in Germany), use Wikipedia more often for intrinsic learning. 

Following the interesting correlations with the HDI, we also tried other socio-economic indicators such as the \emph{GDP per capita}, share of the adult population with \emph{secondary education} (also shown in \Tabref{tab:corr_hdi}), the \emph{Gini coefficient}  of the income distribution and \emph{Internet availability} (not shown due to space limitations). These indicators correlate strongly with HDI and each other. Thus, they also exhibit very similar correlations with the survey response. Since the correlations are somewhat weaker than for the HDI, we cannot identify a single component of the compound HDI measure that appears most relevant for the correlation with Wikipedia readers' intent.

\begin{figure}[t!]
	\centering
	\subfloat[HDI \vs intrinsic learning]{\includegraphics[width=0.5\columnwidth]{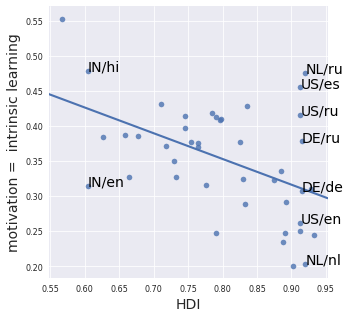}
	\label{fig:hdi_vs_intrinsic}}
	\subfloat[HDI \vs fact checking]{\includegraphics[width=0.5\columnwidth]{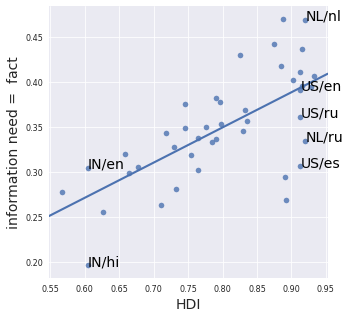}
	\label{fig:shdi_vs_fact}}
	\caption{Correlation between the Human Development Index (HDI) and survey responses. Labels mark data points for countries with multiple language editions, capital letters encode the country, lower case letters, the Wikipedia edition.}
	\label{fig:hdi_vs_intrinsic}
\end{figure}

\para{Human Development Index \vs topics.}
Next, we investigate if the differences in use case prevalences in countries with diverse socio-economic preconditions also manifest themselves in different topics being viewed.
For this purpose, we focus on the Spanish Wikipedia edition, since it is viewed from many countries with diverse HDI scores.
As this analysis does not directly require survey data, we take all Wikipedia readers of our random sample of users and group them by the country their requests came from.
By doing so, we obtain data from 24 countries with more than 500 users each.
We then compute, for all these countries, the viewing likelihood of each of the 20 topics computed via Latent Dirichlet Allocation (\Secref{sec:auxdata}).

\begin{table}[b]
\small
\setlength\tabcolsep{1mm}
\caption{Correlation between survey responses with socio-economic and cultural indicators on a country level, \emph{i.e.}, the Human Development Index (HDI), the GDP per capita, the share of adult population with secondary education, as well as Hofstede's Long-Term Orientation (LTO) and Individualism (IDV) dimensions. The table reports the Spearman correlation coefficient, asterisks indicate the $p$-value of the coefficient under the null hypothesis of independence of the data points (***$<0.001$,**$<0.01$, *$<0.05$). The table is sorted by correlation with HDI.}
\label{tab:corr_hdi}
\begin{tabular}{ll|rrr|rr}
\toprule
&Response &  HDI\parbox{3mm} {~} &  \parbox{7mm}{\centering GDP p.~cap.}\parbox{3mm} {~} &  \parbox{7mm}{\centering Second. educ.}\parbox{3mm} {~} &  LTO \parbox{3mm} {~}  &  IDV \parbox{3mm} {~} \\
\midrule
\multirow{8}{*}{\rotatebox[origin=c]{90}{Motivation}}

& media              &  0.63\parbox{3mm} {***} &     0.58\parbox{3mm} {***}  & 0.42\parbox{3mm} {~} & 0.39\parbox{3mm} {~} & 0.63\parbox{3mm} {***} \\
& work/school        & -0.55\parbox{3mm} {**} &     -0.54\parbox{3mm} {**} & -0.40\parbox{3mm} {~} & -0.37\parbox{3mm} {~} & -0.77\parbox{3mm} {***}\\
& current event      & -0.45\parbox{3mm} {*} &     -0.48\parbox{3mm} {*} & -0.20\parbox{3mm} {~} & 0.13\parbox{3mm} {~} & -0.38\parbox{3mm} {~}\\
& intrinsic learning & -0.40\parbox{3mm} {~} &     -0.43\parbox{3mm} {~} & -0.20\parbox{3mm} {~} & 0.00\parbox{3mm} {~} & -0.26\parbox{3mm} {~}\\
& personal decision  & -0.28\parbox{3mm} {~} &     -0.32\parbox{3mm} {~} & -0.08\parbox{3mm} {~} & 0.31\parbox{3mm} {~} & -0.14\parbox{3mm} {~}\\
& other              & 0.26\parbox{3mm} {~}  &   0.35\parbox{3mm} {~} & -0.08\parbox{3mm} {~} & -0.37\parbox{3mm} {~} &0.04\parbox{3mm} {~}\\
& bored/random       &  0.21\parbox{3mm} {~} &     0.25\parbox{3mm} {~} & -0.02\parbox{3mm} {~} &  -0.17\parbox{3mm} {~} & 0.17\parbox{3mm} {~}\\
& conversation       & -0.07\parbox{3mm} {~} &     -0.12\parbox{3mm} {~} & -0.02\parbox{3mm} {~} & 0.22\parbox{3mm} {~} &0.13\parbox{3mm} {~}\\
\midrule
\multirow{3}{*}{\rotatebox[origin=c]{90}{\parbox{1cm}{\centering info. need}}}
&fact         &  0.66\parbox{3mm} {***} &     0.62\parbox{3mm} {***} & 0.55\parbox{3mm} {**} &0.36\parbox{3mm} {~} & 0.53\parbox{3mm} {*}\\
&in-depth     & -0.60\parbox{3mm} {***} &     -0.57\parbox{3mm} {*} & -0.46\parbox{3mm} {*} & -0.23\parbox{3mm} {~} &-0.43\parbox{3mm} {~}\\
&overview     &  0.25\parbox{3mm} {~} &     0.27\parbox{3mm} {~} & 0.11\parbox{3mm} {~} & -0.13\parbox{3mm} {~} &0.06\parbox{3mm} {~}\\
\midrule
\multirow{2}{*}{\rotatebox[origin=c]{90}{\parbox{0.7cm}{\centering prior knowl.}}}

&familiar      &   0.44\parbox{3mm} {*} &     0.39\parbox{3mm} {~} & 0.47\parbox{3mm} {*} & 0.27\parbox{3mm} {~} &0.42\parbox{3mm} {~}\\
&unfamiliar    & -0.44\parbox{3mm} {*} &     -0.39\parbox{3mm} {~} & -0.47\parbox{3mm} {*} & -0.27\parbox{3mm} {~} &-0.42\parbox{3mm} {~}\\
\bottomrule
\end{tabular}
\end{table}

We observe that several topics exhibit significant correlations with the HDI of the reader's country.
The topics \emph{Math, Physics \& Technology} (Spearman's $r_s = -0.75$, Bonferroni-corrected $p$-value $p < 0.001$), \emph{Research \& Education} ($r_s = -0.73$, $p < 0.001$), and  \emph{Medicine \& Biology} ($r_s = -0.71$, $p \approx 0.002$) show the strongest negative correlations, \emph{i.e.}, these topics are more often viewed in less developed countries. 
By contrast, topics such as \emph{Media Culture} ($r_s =  0.71$, $p \approx 0.002$)  and \emph{Numbers, Lists \& Sports} ($r_s =  0.60$, $p \approx 0.03$) show a significant positive correlation, \emph{i.e.}, articles on those topics are more commonly requested by readers in industrialized countries.
Overall, we observe the tendency that entertainment-oriented topics are more popular in countries with a high HDI, while science\hyp oriented topics are more prevalent in less developed countries.

For the English Wikipedia, we can also find differences between topics across the countries, but the correlations show a less clear picture, partly because many topics obtained from LDA are focused on articles with a specific regional background. 
For example, the topic most strongly correlated with the HDI  is \emph{Geography/US} ($r_s = 0.57$, $p < 0.001$), while the strongest negative correlation is for the topic \emph{Asia} ($r_s = -0.53, p = 0.004$).

\para{Correlation with cultural dimensions.}
To investigate cultural influences on reading behavior, we use Hofstede's cultural dimensions, a well-established and comprehensive framework for characterizing national cultures~\cite{hofstede1984hofstede,hofstede1991cultures}. Hofstede's framework utilizes six dimensions: \emph{Power Distance}, \emph{Individualism}, \emph{Uncertainty Avoidance}, \emph{Masculinity}, \emph{Long-Term Orientation}, and \emph{Indulgence}.

Correlating country-level measures for these dimensions with the survey responses (\Tabref{tab:corr_hdi}, two rightmost columns) shows primarily weak to moderate correlations.
An exception is the \emph{Individualism} (IDV) dimension, for which we observe a clear association between Wikipedia visits motivated by media or work/school: in countries with more collectivist societies (low individualism score) people are less likely to be motivated to visit Wikipedia by media, while using Wikipedia motivated by work or school related tasks is significantly more likely.
\subsection{Summary of Results}
\label{sec:summary}

\para{Survey responses.}
We have shown that Wikipedia is read for a variety of use cases across the 14 Wikipedia languages, and no one use case dominates the others.
Moreover,
we have observed that the prevalence of Wikipedia use cases differs significantly across languages. More specifically, we observe that \emph{intrinsic learning} is the most commonly reported motivation for visiting Wikipedia, except in a minority of languages (including English) where \emph{media} is most common. We also show that the motivations \emph{work\slash school} and \emph{bored\slash random} have the highest prevalence discrepancies.
We observe that information need and prior knowledge vary to a great extent across languages, with the reported \emph{in-depth} reading ranging from 21\% to over 60\% and \emph{familiarity} from less than 40\% to more than 70\%.

Through the analysis of the survey results we show that the English Wikipedia---the sole focus of many Wikipedia studies--- is not representative of all Wikipedia languages.
Rather, it can be considered an outlier with regard to several aspects.
Finally, the survey results for the English Wikipedia line up well with the previous study, providing evidence for the robustness of results over time.

\para{Usage patterns.} 
By connecting survey responses to request logs we have identified several usage patterns in the logs that can be consistently associated with certain Wikipedia use cases across languages. Specifically, motivation \emph{bored\slash random} can be linked to certain patterns including long sessions with rapid requests and internal browsing, while  \emph{work\slash school} can be linked to slow requests to desktop versions of Wikipedia.
We also observe that not all patterns discovered for English Wikipedia use cases \cite{singer2017we} hold for the other Wikipedia languages. Furthermore, the different use cases in the languages alone are not sufficient to explain the differences in usage patterns across languages.

\para{Country-level statistics. }
We find significant correlations between the Human Development Index of a country and the prevalence of Wikipedia use cases reported from there.
In particular, less developed countries are more likely to read Wikipedia \emph{in depth} and be motivated by \emph{work\slash school} or \emph{intrinsic learning}. In industrialized countries, Wikipedia readers are more often checking \emph{facts} and are triggered to visit Wikipedia by \emph{media}.
Socio-economic differences also show in different topics being viewed:  science\hyp oriented topics
are more important for less developed countries, while entertainment-oriented topics are more common in industrialized countries. 
Cultural factors as measured by Hofstede's cultural dimension, with the exception of \emph{Individualism,} seem to play a lesser role.

\section{Discussion}
\label{sec:discussion}
In this section, we present the implications of this study, future directions, and methodological limitations.

\subsection{Implications and Future Directions}

\para{Beyond English Wikipedia.}
A tremendous amount of research and development on Wikipedia has been focused on, or informed by, English Wikipedia. This study sheds light on the importance of breaking this cycle and acknowledging that English Wikipedia is not representative of all Wikipedia languages, and in several aspects is an outlier. Wikipedia's endeavor towards knowledge equity\footnote{\url{https://meta.wikimedia.org/wiki/Strategy/Wikimedia_movement/2017/Direction\#Our_strategic_direction:_Service_and_Equity}} requires a deeper and better understanding of the differences between Wikipedia languages.
This work should
be expanded to enhance our understanding of the access to, and production of, knowledge in Wikipedia. Future studies can investigate the socio-economic factors at a finer granularity than the country level, include demographic information, and attempt to characterize potential Wikipedia readers.

\para{Global \vs\ local solutions.} One of the findings of the current study is that, except for a few general patterns, the patterns that describe readers' use cases of Wikipedia differ across languages. This indicates that one-size-fits-all solutions may not work across languages, and a combination of global and local solutions may be needed to satisfy the needs of Wikipedia readers. Future research should focus on scaling locally aware solutions across many languages. 

\para{Within-session language switching.} 
Our preliminary analysis shows that on average roughly 20\% of reader sessions involve the reader switching from one Wikipedia language to another. Future work can investigate circumstances under which users switch from one language to another, which can in turn inform the prioritization of content creation across Wikipedia languages.

%

\subsection{Methodological Limitations}
\label{sec:limitations}
\para{User identification.} 
Wikipedia does not require users to log in, nor does it use cookies in webrequest logs to maintain a notion of unique clients. Therefore, we rely on a combination of IP addresses and user agents to approximate unique devices (\cf\ Singer \etal\ \cite{singer2017we} for an in-depth discussion of this approximation and its limitations).

\para{Survey-response bias.}
Not all users are equally likely to participate in a voluntary survey, \emph{i.e.}, some groups will be overrepresented in the survey responses. To tackle this issue, we reweighted the responses based on features from the server logs by inverse propensity weighting. However, if other covariates (\emph{e.g.}, age or gender) that are not explicit in the server logs influence the responses, these might skew the results.

\para{Translation bias.} Multilingual surveys suffer from differences in the translations of survey questions and answer options. For this study, translations were carefully done by Wikipedia editors who are native speakers of the languages in this study. Translated content was then checked word by word in online meetings between the translator(s) and one of the study authors. Even with a process such as above, we cannot rule out different nuances and connotations between the languages that may not have been captured as part of the translations.

\para{Social desirability.} Survey responses are commonly subject to \emph{social desirability bias}~\cite{nederhof1985methods}, \emph{i.e.}, participants are more likely to reply with options that are viewed in a positive light in their society. Even though the questions in our survey are of non-sensitive nature and the survey is done anonymously, social desirability could still influence our results; \emph{e.g.}, browsing Wikipedia out of boredom could be seen as negative in some cultures and therefore might be picked less often by survey participants. As the effect of such a bias could be different in different societies,
this might skew comparisons between languages.

\section{Conclusions}
\label{sec:conclusions}
In this work, we study why users from around the world read Wikipedia. Through a large-scale survey with more than 210,000 responses across 14 Wikipedia languages we highlight key commonalities and differences in Wikipedia use cases across these languages. Combining the survey responses with webrequest logs as well as country level socio-economic statistics allows us to characterize Wikipedia use cases across languages with behavioral patterns and socio-economics indicators. The outcomes of this study provide a deeper understanding of Wikipedia readership in a wide range of languages, which is important for Wikipedia editors, developers, and the reusers of Wikipedia content.


%
\setcode{utf8}
\section*{Acknowledgements}
We thank the following Wikipedia users for translating the survey to their languages and acting as our points of contact in their language communities:  \<عباد\_ديرانية >, Amire80, Antanana, AWossink, Hasive, Kaganer, Lyzzy, Rasco, Satdeep\_Gill, Shangkuanlc, Strainu, Tgr, Whym. 
We thank Philipp Singer and Ellery Wulczyn for contributing their code of the predecessor study, Markus Strohmaier for helpful discussions, and Bahodir Mansurov for running the surveys on Wikipedia.


%


\begingroup
\sloppy
\balance
\bibliographystyle{ACM-Reference-Format}
\bibliography{bib}  
%


\end{document}